\documentclass[12pt]{iopart}

\usepackage{sidecap}
\usepackage{bm}

\usepackage{amsfonts}
\usepackage{color}

\usepackage{graphicx}
\usepackage[english]{babel}

\definecolor{dred}{rgb}{0.6,0.1,0}

\newcommand{\wcmq}{\mathrm{W}/\mathrm{cm}^2}

\newcommand{\omlas}{\omega_\mathrm{las}}
\newcommand{\llas}{I_\mathrm{las}}

\newcommand{\bohr}{\mathrm{a}_0}

\newcommand{\NaCl}{Na$^+_9$ }
\newcommand{\ArCl}{Ar }
\definecolor{dgreen}{rgb}{0,0.5,0}

\begin{document}

\title[The impact of the carrier envelope phase ...]
{The impact of the carrier envelope phase -- dependence on system and
  laser parameters }

\author{P.-G. Reinhard$^{1}$, E. Suraud$^2$, C. Meier$^{3}$}

%Neil Scriven\dag\ and Romneya Robertson\ddag
%\footnote[3]{To
%whom correspondence should be addressed (nester@thsun1.jinr.ru)}
%}
\address{$^1$ Institut f\"ur Theoretische Physik,
 Universit\"at Erlangen, D-91058, Erlangen, Germany}
%\\
\address{$^2$ Laboratoire de Physique Th\'eorique, Universit\'e de
  Toulouse, CNRS, UPS, France}
%\\
\address{$^3$ Laboratoire Collisions-Agr\'egats-R\'eactivit\'e, 
Universit\'e de Toulouse, CNRS, UPS, France}

\ead{paul-gerhard.reinhard@physik.uni-erlangen.de}

\begin{abstract}
We investigate, from a theoretical perspective, photo-emission of
electrons induced by ultra-short infrared pulses covering only a few
photon cycles. In particular, we investigate the impact of the
Carrier-Envelope Phase (CEP) of the laser pulse which plays an
increasingly large role for decreasing pulse length. As key observable
we look at the asymmetry of the angular distribution as function of
kinetic energy of the emitted electrons. The focus of the present
study lies on the system dependence of the reaction. To this end, we
study two very different systems in comparison, an Ar atom and the
\NaCl cluster. The study employs a fully quantum-mechanical
description of electron dynamics at the level of Time-Dependent
Density Functional Theory (TDDFT). We find a sensitive dependence on
the system which can be related to the different spectral response
properties. Results can be understood from an interplay 
of the ponderomotive motion driven by the external photon field and
dynamical polarization of the system. %The latter
\end{abstract}



\submitto{\JPB}


%\maketitle

\section{Introduction}
\label{sec:intro}

The enormous progress in laser technology has meanwhile reached the
regime of extremely short pulses which cover only a few optical cycles
\cite{Nis97,Bra00}. This opens the door to a great variety of new
investigations such as time-resolved measurement of electronic
processes or precision control of chemical reactions see, e.g.,
\cite{Bal03,Kli06,Mil06,Hae11,Liu11,Kru12a,Xie12}.  One of the main
new aspects of such few-cycle pulses is that the Carrier Envelope
Phase (CEP), i.e. the relative phase between the optical carrier wave
and the pulse envelope, acquires a sensitive influence on the pulse
shape which, in turn, can have a strong impact on all laser induced
reactions \cite{Tem99,PauG01b,Che05}.  This looks at first glance as a
complication because there is one more laser parameter to be taken
care of. But it can be turned into an advantage by potentially
controlling electronic reactions with dedicated variations of the CEP.
A particularly interesting example is found in photo-electron emission
where the CEP allows to control the forward-backward (also called
``right-left'') asymmetry in the photo-electron spectra
(PES)~\cite{Pau03,Cor93}. Theoretical calculations have shown that the
(high energy) parts of PES coming from electron recollisions are more
sensitive to CEP than the (low energy) parts from directly emitted
electrons \cite{Mil03,Che05,Ton06,Qin08,Sua15}.  Experimentally, the
dependence of high-energy PES on the CEP has been explored for
atoms~\cite{Pau03,Lin05,Kli08b}, dimer molecules~\cite{Gaz11}, and
nano-tips \cite{Kru11,Par12}, providing challenging motivations for
the theoretical investigations.

In a recent work, the Angular-Resolved PES (ARPES) of
C$_{60}$ were calculated as a function of the Carrier Envelope Phase (CEP)
using a fully quantum-mechanical approach based on Time Dependent
Density Functional Theory  (TDDFT) and looking in particular at the asymmetry
of the angular distribution for different kinetic energies of the
emitted electrons \cite{Gao17a}.  These theoretical simulations
allowed to reproduce the experimentally found dependencies on laser
parameters, particularly on pulse length and CEP.  The next question
which comes up naturally is to which extend the results depend on the
system which is irradiated. To answer this question, we have extended
the previous calculations to include different atomic systems, notably
atomic Argon and \NaCl clusters.  These systems were chosen, since
they have about the same complexity (8 active electrons) but differ
drastically in their electronic response: the photo-excitation
spectrum of \ArCl has a discrete peak at about 15 eV followed by a
widespread shoulder above while the spectrum of \NaCl is dominated by
a strong surface plasmon resonance at about 2.7 eV. It is obvious that
an infrared (IR) laser is far closer to resonance for \NaCl and this
should have consequences. To investigate this in detail is the aim of
the present work.

The paper is organized as follows: In section \ref{sec:formal}, we
briefly review the theoretical framework. In section \ref{sec:result}, 
we present and discuss the results.

\section{Formal framework}
\label{sec:formal}

%%\subsection{Modeling}
%%\label{subsec:fram}

We describe the electronic dynamics by Time-Dependent Density
Functional Theory (TDDFT) at the level of the Time-Dependent Local
Density Approximation (TDLDA)~\cite{Dre90} using the
exchange-correlation functional from~\cite{Per92}.  For an appropriate
modeling of electron emission, we augment the TDLDA by a Self-Interaction
Correction (SIC) \cite{PeZ81}. As a full SIC treatment is
computationally very demanding, \cite{Mes08b}, we use it in a
simplified, but reliable and efficient version as an Average Density
SIC (ADSIC)~\cite{Leg02}. The ADSIC is able to put the single-particle
energies into the correct relation to the continuum threshold such
that the ionization potential (IP) is properly reproduced in a great
variety of systems~\cite{Klu13a}. A correct description of the IP is
particularly important for the analysis  of photoemission, the more so
in connection with strong IR fields because here electrons in the
highest occupied molecular orbitals dominate ~\cite{Mul96}.

The ionic background is described by soft local pseudopotentials, for
\NaCl of Gaussian type \cite{Kue00} and for Ar using the functional
form of \cite{Goe96}. The ions of \NaCl are kept frozen at the
ground-state structure during the dynamical calculations. This is a
legitimate approximation in view of the short laser pulses considered
here.

The external, linearly polarized laser pulse is modeled within dipole
approximation as the potential
%the potential (in atomic units) laser
%pulse with the interaction with the laser field is given by
\numparts
\begin{eqnarray} 
  v_{\rm las}(\mathbf r,t) 
  & = &  
  -E(t) \, \mathbf{r} \cdot \mathbf{e}_z
\end{eqnarray}
with  the polarization vector along the $z$-axis $\mathbf{e}_z$ and
with the electric field
\begin{eqnarray}
\label{eq:laser}
  E(t) 
  & = & 
  E_0 \cos^2\left(\frac{\pi t}{T_\mathrm{pulse}}\right) 
  \cos(\omlas t+\phi_\mathrm{CEP}) 
  \;\mbox{for}\;
  -\frac{T_\mathrm{pulse}}{2} \leq t \leq \frac{T_\mathrm{pulse}}{2}
  \;.
\end{eqnarray}
\endnumparts 
Here, $E_0$ denotes the peak electric field, $\omlas$ the carrier
frequency, and $T_\mathrm{pulse}$ the total pulse duration.  The CEP
is comprised in the parameter $\phi_\mathrm{CEP}$ which defines the
phase between a maximum of oscillations with frequency $\omlas$ and
the maximum of the $\cos^2$ envelope. The minus sign in the
  definition of the potential guarantees that the driving force of the
  electrical field points in forward direction (positive $z$) for
  positive field $E(t)$.  In what follows, we use laser parameters
close to those in recent experiments~\cite{Li15}: frequency
$\omlas=1.72$ eV (a wavelength of 720 nm), intensity
$I=6\times10^{13}~\wcmq$, corresponding to field amplitude $E_0=1.1$
eV/a$_0$, and total duration $T_\mathrm{pulse}$=4 fs, 6 fs, and 8 fs,
corresponding to 1.7, 2.5, and 3.3 optical cycles (1 optical cycle
$/tau_\mathrm{IR}$ = 2.4 fs). Note that these laser parameters are
associated with a ponderomotive energy $U_p=2.9$ eV.
\begin{SCfigure}%[0.5]
{\includegraphics[width=0.5\linewidth]{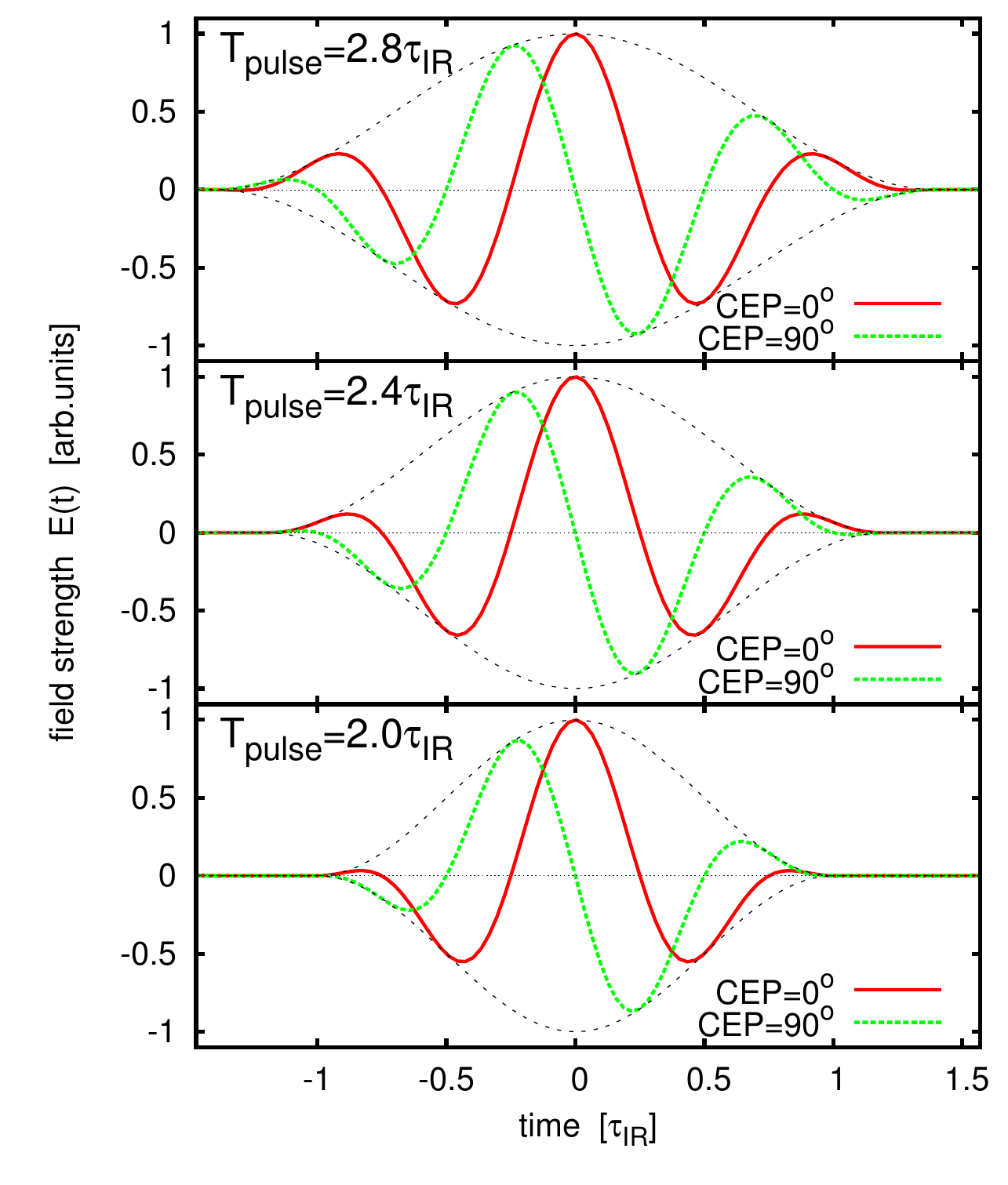}}
 \caption{\label{fig:pulses} Examples of IR pulses for various pulse
   lengths $T_\mathrm{pulse}$ and two CEP as indicated. The dashed
     lines indicate the pulse envelope which does not depend on the CEP.}
\end{SCfigure}
Figure~\ref{fig:pulses} illustrates the temporal part of the laser
field $\propto E(t)$ for the three $T_\mathrm{pulse}$ under
consideration, each one for the CEP at $0^{\circ}$ and
$90^{\circ}$.
For $\phi_\mathrm{CEP}=0^{\circ}$, the center of the envelope (dashes)
coincides with a maximum of the oscillations, while for
$\phi_\mathrm{CEP}=90^{\circ}$, it is shifted to match with the
nodal points of the electric field. Clearly, we see a substantial
change of the electric field due to the different CEP in each case. 

\subsection{Numerical details}
\label{subsec:numerics}

The TDLDA-ADSIC equations are solved numerically on a cylindrical grid
in coordinate space~\cite{Mon95a}. The static iterations towards
the electronic ground state are done with the damped gradient
method~\cite{Rei82} and the time evolution employs the time-splitting
technique~\cite{Fei82}. For details of the numerical method
see~\cite{Cal00,Rei03a,Wop15}.
We use a numerical box which extends to $\pm 250~\bohr$
in $z$ direction (along the laser polarization) and $250~\bohr$ orthogonal to it
(radial $r$ coordinate), with a grid spacing of $0.5~\bohr$ in both
directions.
Time propagation is followed up to after 44 fs with a small time step
of 1--5 attoseconds. Box size and time span are sufficiently large to
track completely the rescattering of electrons in the laser field
(ponderomotive motion). To account for ionization, absorbing
boundary conditions are implemented using a mask function~\cite{PGR06}.
The absorbing margin extends over 32 $\bohr$ at each side of the grid.

The central observable of electron emission in our analysis is the
Angle-Resolved Photo-Electron Spectrum (ARPES), i.e., the yield of
emitted electrons [$\mathcal{Y}(E_\mathrm{kin},\theta)$] as function
of kinetic energy $E_\mathrm{kin}$ and emission angle $\theta$. We
calculate an ARPES by recording at each time step the single-electron
wave functions \{$\psi_j(t,\mathbf{r}_{\mathcal{M}})$, $j=1, \ldots,
N_\mathrm{el}$\} at selected measuring points
$\mathbf{r}_{\mathcal{M}}$ near the absorbing layer and finally
transforming this information from time- to frequency-domain,
see~\cite{Poh01,DeG12,Din13,Dau16}. Finally, the ARPES is written as
\begin{equation}
  \mathcal{Y}(E_\mathrm{kin},\theta) 
  \propto 
  \sum_{j=1}^{N_{\mathrm{el}}}
 | \widetilde{\psi_{j}}(E_\mathrm{kin},\mathbf r_{\mathcal M})
 |^2 
\label{eq:pes}
\end{equation}
where $\widetilde{\psi_{j}}$ are the transformed wave functions in
energy domain. In case of strong fields, as we encounter here, the
$\widetilde{\psi_{j}}$ are augmented by a phase factor accounting for
the ponderomotive motion, for technical details see \cite{Din13}.  The
angle $\theta$ is defined with respect to $\mathbf{e}_z$,
i.e. $\theta=0^{\circ}$ means electronic emission in the direction of
$\mathbf{e}_z$.  A detailed ARPES analysis requires a fine resolution
for Fourier transformation and emission angles. To that end, we use an
increment of 0.04 eV in energy and $1^{\circ}$ opening angle for the
angular bins.

It is cumbersome and confusing to check the full ARPES for all
variations of laser parameters. It turns out that a key feature is the
asymmetry of the angular distribution
\numparts
\label{eq:eta}
\begin{eqnarray}
  \eta
  &=&
  \frac{{Y}_+(E_\mathrm{kin},\Theta)
           -{Y}_-(E_\mathrm{kin},\Theta)}
       {{Y}_+(E_\mathrm{kin},\Theta)
           +{Y}_-(E_\mathrm{kin},\Theta)}
  \quad,
\\
  {Y}_+(E_\mathrm{kin},\Theta)
  &=&
  \int_0^\Theta\!d\theta\,\mathcal{Y}(E_\mathrm{kin},\theta) 
\\
  {Y}_-(E_\mathrm{kin},\Theta)
  &=&
  \int_{180^{\circ}-\Theta}^{180^{\circ}}\!d\theta\,\mathcal{Y}(E_\mathrm{kin},\theta) 
\end{eqnarray}
\endnumparts 
which is a function of kinetic energy $E_\mathrm{kin}$ and opening
angle $\Theta$ over which angular averaging is done.  In most
experiments, photoelectron yields are collected in a cone angle of
$\Theta=15^{\circ}$. Thus we are using this value throughout and will
not indicate this dependence any more. Of course, this $\eta$ depends
on more parameters, particularly on all laser parameters. We will in the
following look particularly at the dependence on the CEP.

In some cases, we compress the information by integrating over certain
energy bands. This yields  a more compact asymmetry parameter
%\numparts
\begin{eqnarray}
  \eta[E_1:E_2]
  &=&
  \frac{\tilde{Y}_+[E_1:E_2]
           -\tilde{Y}_-[E_1:E_2]}
       {\tilde{Y}_+[E_1:E_2]
           +\tilde{Y}_-[E_1:E_2]}
 \quad,
\label{eq:etaint1}
\\
  \tilde{Y}_\pm[E_1:E_2]
  &=&
  \int_{E_1}^{E_2}\!dE_\mathrm{kin}\,{Y}_\pm(E_\mathrm{kin},15^{\circ})
  \quad,
\label{eq:etaint2}
\end{eqnarray}
%\endnumparts 
where we use the same symbol $\eta$ and distinguish by indicating its
dependencies.

%\clearpage

\section{Results}
\label{sec:result}

\subsection{Introductory example}
\label{sec:introex}

\begin{figure}%[0.5]
\centerline{\includegraphics[width=0.6\linewidth]{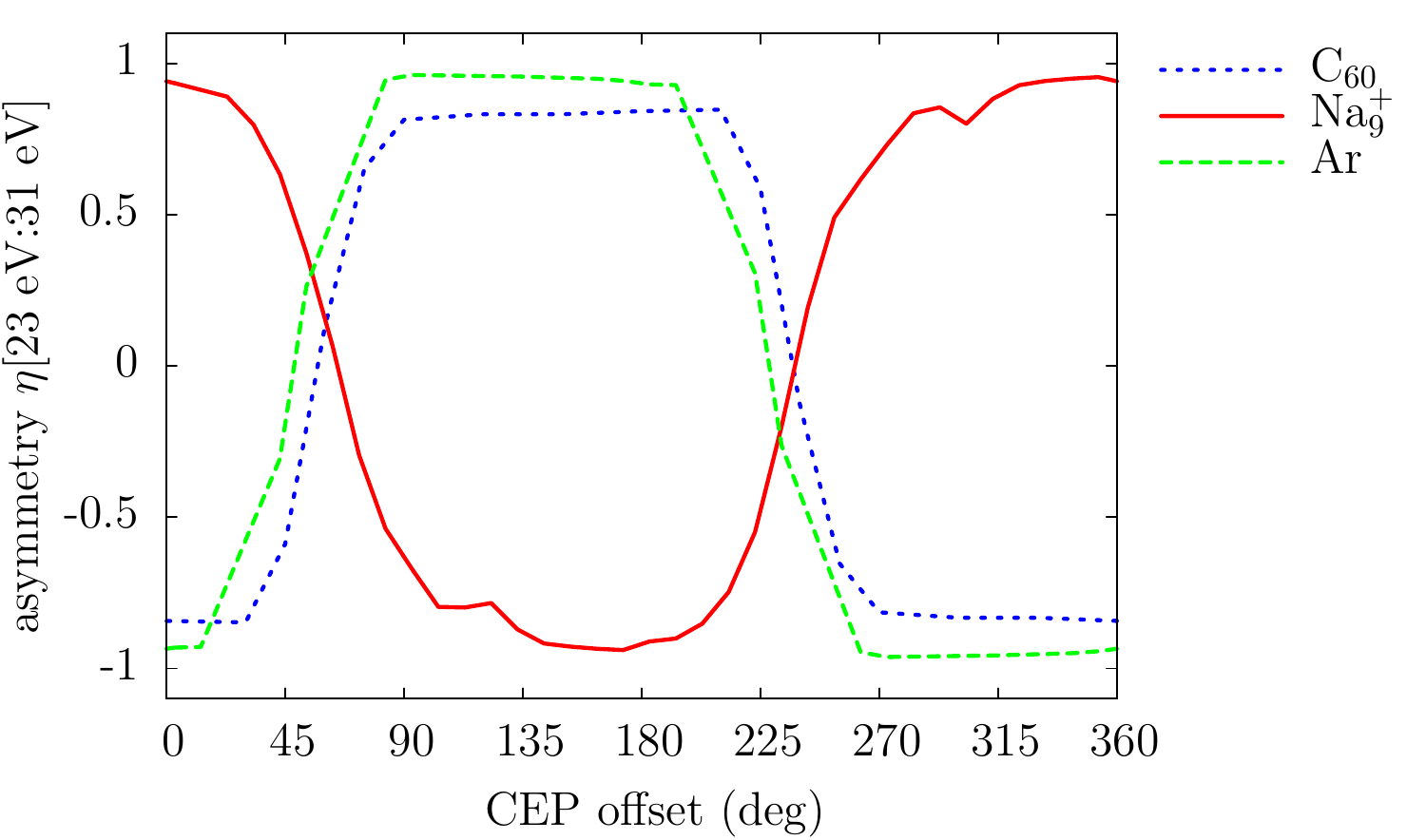}}
 \caption{\label{fig:old} Integrated asymmetry parameter
   $\eta[23\mathrm{eV}:31\mathrm{eV}]$ for three different systems:
   C$_{60}$ cluster, Ar atom, and Na$_9^+$ cluster.  Laser parameters
   were: frequency $\omlas=1.72$ eV, intensity
   $\llas=6\times10^{13}~\wcmq$ corresponding to field strength
   $E_0=1.1$ eV/a$_0$, and pulse length $T_\mathrm{pulse}$=6 fs
   (corresponding to 2.5 IR cycles).
 }
\end{figure}
For a first comparison of systems, we chose the integrated asymmetry
Eq. (\ref{eq:etaint1}) versus CEP as simple signal. Results for three
different systems are compared in figure \ref{fig:old}: the C$_{60}$
cluster to establish the link to the previous study \cite{Gao17a}, the
Ar atom, and the \NaCl cluster. The first two samples, C$_{60}$ and
\ArCl, show a very similar behavior and one is tempted to expect a
generic signal formed only by laser properties. However, the \NaCl
cluster yields a CEP dependence with exactly opposite pattern which
clearly demonstrates a strong impact of the electronic system.  The
difference between Ar and \NaCl suggests naturally a relation to their
much different spectral response. This assumption is, in fact,
confirmed by the case of C$_{60}$. Although this large cluster is
structurally much different from the Ar atom the spectral
relations are similar to Ar in that the dipole strength  lies far
above the IR frequency and is rather fragmented. We shall thus continue
our investigations  keeping only  \ArCl and \NaCl for the sake of simplicity. 

\subsection{Comparison of  Ar atom and Na$_9^+$ cluster}

\begin{figure}
 \centerline{\includegraphics[width=0.72\linewidth]{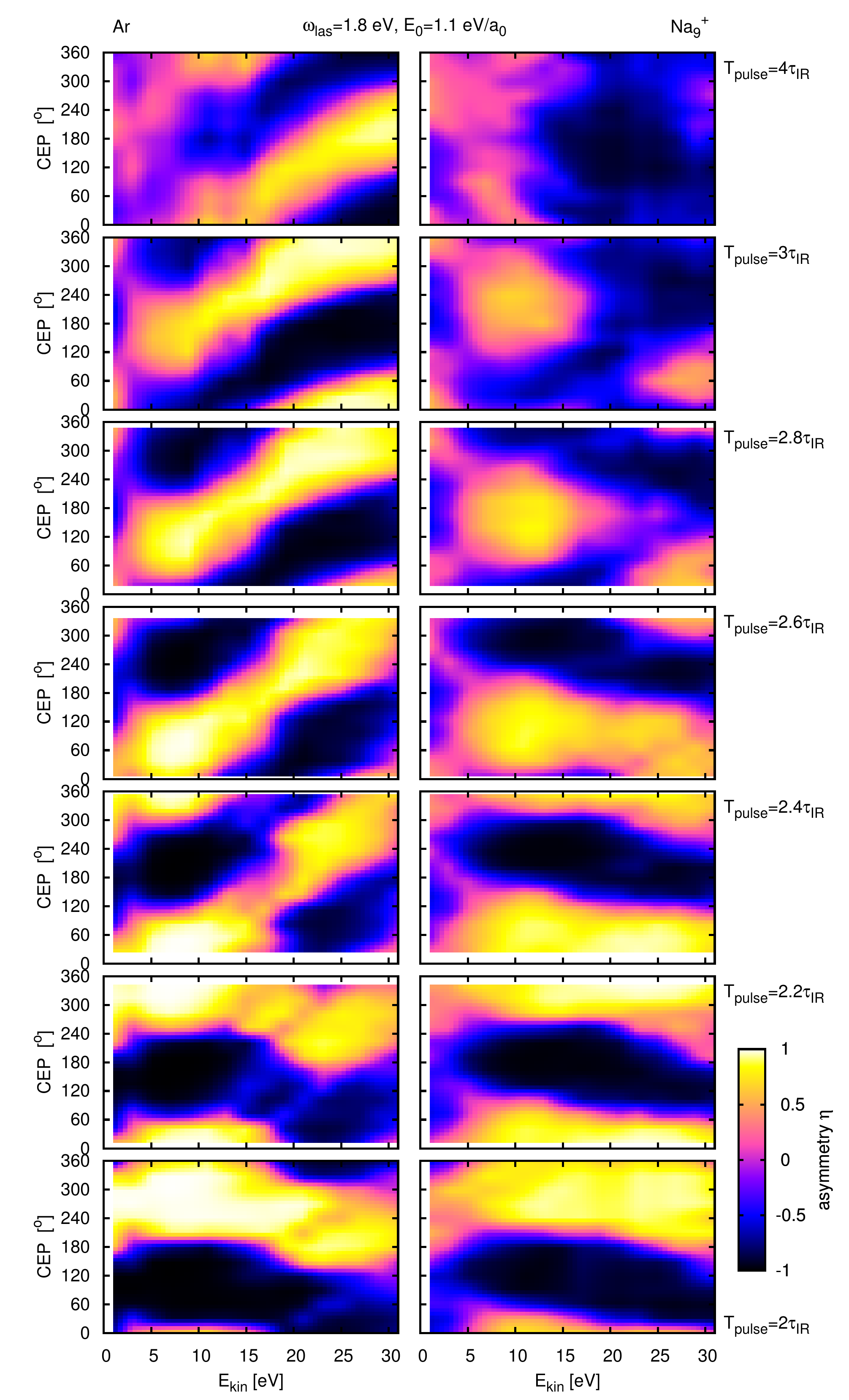}}
 \caption{\label{fig:compare-asym-bins} Asymmetry $\eta$ in the
   plane of kinetic energy and CEP for $\omlas=1.8$ eV, $E_0=1.1$
   eV/a$_0$, and different pulse lengths as indicated. }
\end{figure}
In this section, the simultaneous dependence of the asymmetry
parameter $\eta$ on the CEP and the kinetic energy $E_\mathrm{kin}$ of
emitted electrons is analyzed in detail.  The results of this 3D
information are presented as a color map.  Figure
\ref{fig:compare-asym-bins} shows a series of results for
different pulse lengths $T_\mathrm{pulse}$.
%
%\begin{figure}
% \centerline{\includegraphics[width=0.7\linewidth]{compare-asym-bins-high.pdf}}
% \caption{\label{fig:compare-asym-bins-high}
%As figure \ref{fig:compare-asym-bins-low} for a couple of longer pulses.
% }
%\end{figure}
%
Let us first look at extremely short pulses,
$T_\mathrm{pulse}=2\tau_\mathrm{IR}$, i.e. practically less than two
optical cycles because of the the profile shape (see
Fig. \ref{fig:pulses}). Here, the CEP dependence of the asymmetry is
very similar for both systems (\ArCl and \NaCl) and the asymmetry is
more or less the same over the whole range of $E_\mathrm{kin}$. The
independence from the system is plausible because very short pulses
give the system little chance to develop its dynamical response such
that the shape of the external laser pulse dominates the process and
this is the same in all systems.  The independence from the CEP is
also plausible because too short pulses have basically one or two
dominating peaks which gives little chances for ponderomotive effects
which otherwise can produce a difference between low-energy and
high-energy emission \cite{Gao17a}. The emission maxima (corresponding
to $\eta \sim \pm 1$) lie at $\phi_\mathrm{CEP}=90^{\circ}$ and
$\phi_\mathrm{CEP}=270^{\circ}$. These are the situations where
exactly two equally strong peaks compete and the preferred direction
of emission coincides with the direction of the second peak.  For
example, the second of the large peaks points to negative $z$ for
$\phi_\mathrm{CEP}=90^{\circ}$, see Fig. \ref{fig:pulses}, and so
backward emission ($\eta\approx -1$) prevails. This agrees with the
analysis of detailed time dependence in section \ref{sec:timedep}
which shows that the emission from earlier peaks is practically pushed
back by the subsequent counter-peaks and that emission from the last
peak experiences least hindrance.

However, when slightly longer pulses are considered, the picture
changes. Directly above $T_\mathrm{pulse}=2\tau_\mathrm{IR}$ comes a
region where pattern change quickly and the difference between Ar and
\NaCl develops.  Already at $T_\mathrm{pulse}=2.4\tau_\mathrm{IR}$ we
can see a striking difference between the patterns for \ArCl and
\NaCl: while in the case of \NaCl, (right panel), the asymmetry has
not changed significantly as compared to
$T_\mathrm{pulse}=2.0\tau_\mathrm{IR}$, the asymmetry has developed a
clear energy dependence for Ar.  In the latter case, the asymmetry in
the low-energy and high-energy ranges show opposite behavior: for
instance, for $\phi_{CEP}=240 \circ$, low energy electrons ($\approx$
5--10 eV) are mainly emitted in forward direction, while high energy
electrons $\approx 22$ eV backward.  The reason for this behavior
lies in the optical properties of the field and the electronic
response: for a pulse duration of $T=2.4 \tau_\mathrm{IR}$, the electric field
explores already a bit more oscillations which enhances the playground
for ponderomotive motion and this, in turn, gives the high-energy
electron a different emission dynamics.  This behavior is similar to
the one found in C$_{60}$ previously and discussed in \cite{Gao17a}.
One then wonders why this does not happen in the same manner for
\NaCl. The key point is here that the external driving field and the
dipole response run out of phase due to the impact of the strong
plasmon resonance. This interferes with the mechanism worked out for Ar
and C$_{60}$. In section \ref{sec:timedep}, the difference in
dynamical behaviors will be analyzed based on the detailed time
evolution of electron density.

Finally, when going to even longer pulses, the asymmetry vanishes
altogether.  The reason for this behavior is obvious. Longer pulses
contain more optical cycles, the individual peaks of the electric
field that point into opposite directions with nearly equal strength
and the CEP does not make much of a difference any more. It is
interesting, however, that the smoothing of asymmetry pattern as a
function of CEP starts for \NaCl already at
$T_\mathrm{pulse}=3\tau_\mathrm{IR}$ while it takes up to much larger
$T_\mathrm{pulse}$ to reach that stage for Ar. The reason is, again,
the different optical response. The interference of external frequency
and resonance frequency in the dipole response of \NaCl perturbs the
emission process and so spoils CEP dependence earlier.

\subsection{Time evolution in detail}
\label{sec:timedep}

In order to get a deeper insight into 
emission direction, we analyze the time evolution of
electron density for a pulse length in the critical regime, 
$T_\mathrm{pulse}=2.6\tau_\mathrm{IR}$, and here for 
$\phi_\mathrm{CEP}=270^{\circ}$ where the difference between Ar and
\NaCl is most marked.

\begin{figure}[b]
 \centerline{\includegraphics[width=0.9\linewidth]{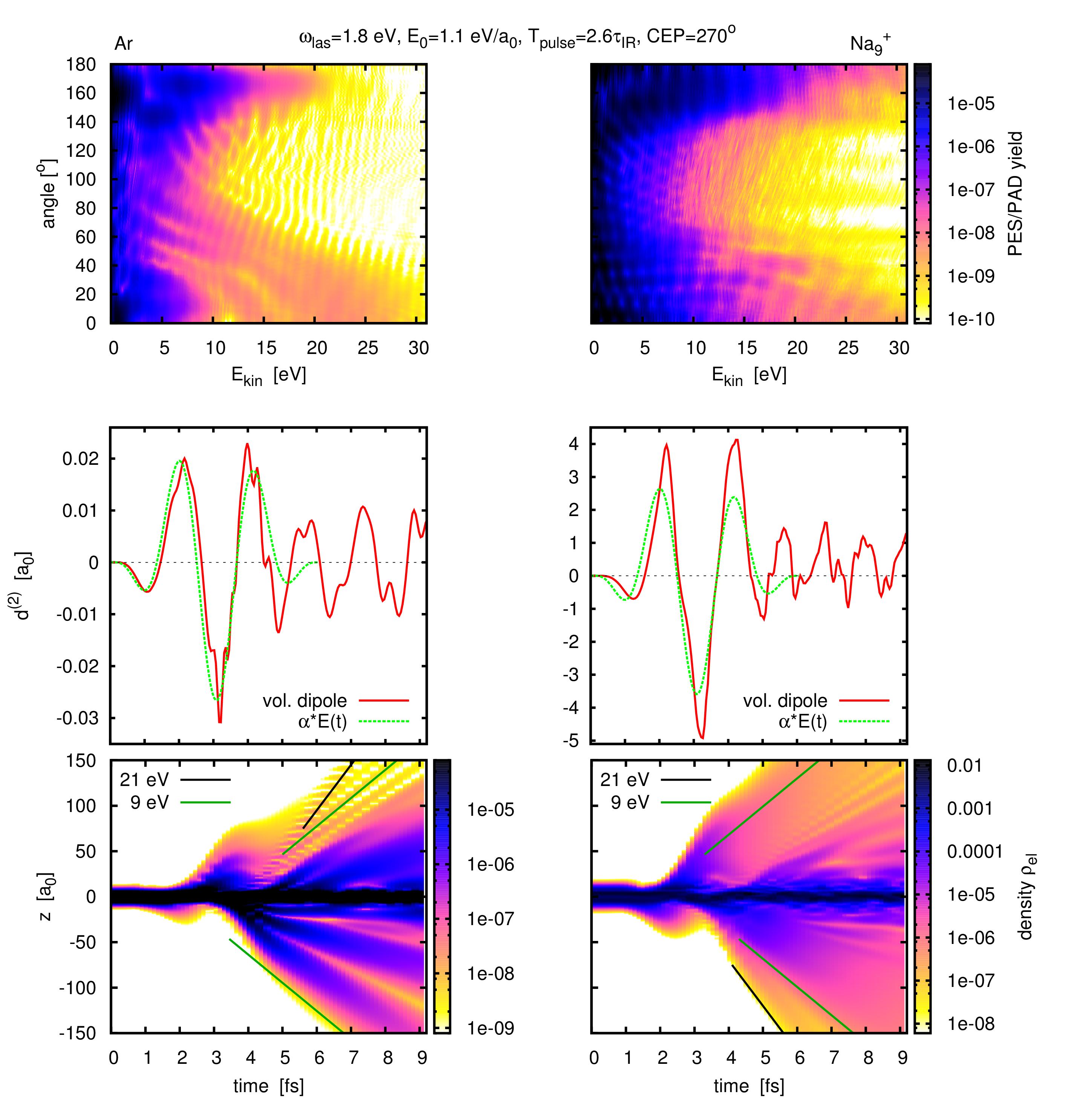}}
 \caption{\label{fig:compare-dens-dipole} Analysis of the case
   $\omlas=1.8$ eV, $E_0=1.1$ eV/a$_0$, pulse length
   $T_\mathrm{pulse}=2.6\tau_\mathrm{IR}$, and CEP=270$^o$ for Ar (left)
   and Na$_9^+$ (right)
 Upper: ARPES yield $\mathcal{Y}$ in the plane of
   kinetic energy $E_\mathrm{kin}$ and emission angle $\theta$.
%\newline
  Middle: external field and volume dipole $d^{(2)}=\int
  d^3r\,z\,\rho^2/\int d^3r\,\rho^2$ as function of time. The external field is scaled
  such that the first maximum matches with the dipole. The scaling
  factor is $\alpha=0.025\mathrm{a}_0^2/$eV for Ar and 
   $\alpha=3.3\mathrm{a}_0^2/$eV for Na$_9^+$.
%\newline
  Lower: Density (in logarithmic color scale) in the plane of
  time and $z$-coordinate.  The black lines indicate the slope of a
  point moving with $E_\mathrm{kin}=21$ eV, typical for the
  high-energy regime, and the green lines
  indicate $E_\mathrm{kin}=9$ eV, typical for the low-energy regime.
  }
\end{figure}
The lower panel of figure \ref{fig:compare-dens-dipole} shows the
densities for Ar and \NaCl as 3D color map plot. This is complemented
by the full ARPES in the upper panel and time evolution of external
field together with volume dipole moment in the middle panel.  The
notion of a ``volume dipole'' requires an explanation. The normal
dipole moment $d=\int d^3r\,z\,\rho$ is blurred by pieces of emitted
electronic density far away from the system, i.e. at large
$z$. However, we are interested on a signal from the bulk density at
the system. In order to focus on the regions of high density, we
consider the $z$ moment of $\rho^2$, i.e.  $d^{(2)}=\int
d^3r\,z\,\rho^2$, which we call volume dipole because it emphasizes
the volume rather than the surface.

The ARPES in the upper panel illustrates how the asymmetry pattern
(see figure \ref{fig:compare-asym-bins}) are generated.  For Ar, we
see a preference of backward emission (angle $180^{\circ}$) at energies
around 5 eV turning to a bias on backward emission for higher energies
$>20$ eV while for Na$_9^+$ backward emission prevails in the whole
range (except for very low energies $<2$ eV), in accordance with the
panel for $T_\mathrm{pulse}=2.6\tau_\mathrm{IR}$ in figure
\ref{fig:compare-asym-bins}.

Dipole and field signals in the middle panel illustrate nicely the
dramatic differences in the systems response. Already the dipole
amplitude for Ar is much smaller than for \NaCl which indicates that
Ar is a much less responsive system. This off-resonant behavior for Ar
is also visible from the fact that the dipole signal follows the
external field in phase. There are small, fast oscillations
super-imposed in the dipole signal which stem from a small excitation
of systems frequencies. The \NaCl cluster behaves much differently.
Here we see already in the first oscillation a large phase shift of
dipole versus external field. The shift changes along the signal. It
seems that the dipole response has a trend to oscillate with the
faster plasmon frequency rather than those of the external field.
Mind that emission is driven by the dipole moment whereas propagation
of the emitted electron follows the external field. These two
ingredients now run at different pace for \NaCl producing a 
subtle interplay of driving forces. The trends thus can go in any
direction and it requires detailed modeling to make predictions for
such highly reactive systems.

Finally, the lower panel of figure \ref{fig:compare-dens-dipole} shows
the time evolution of electron density along $z$ axis (laser
polarization axis). Logarithmic scale is chosen to visualize the outer
tail of the electron distribution which represents electron
emission. In the first 3 fs, we see outbursts of electrons which are,
however, turned back to the center. Consider the first bump toward
negative $z$. This  outflow is triggered by the first peak of $E(t)$,
or $d^{(2)}(T)$ respectively. It moves on until  the opposite peak
in field strengths pushes it back to the center. The same happens with
electrons emitted through the second (positive) peak in direction of
positive $z$. Here the counter-force in negative direction does not
complete its job and part of the backflow merges with the outgoing
wave from the second positive peak. The waves finally reaching the
boundaries stem all from the latest peaks of the pulse which are not
counter-weighted by subsequent opposite peaks. The final outflow mixes
electrons which are directly emitted with those which have gained more
energy by rescattering processes. This is indicated by the lines drawn
into the plot. The black lines visualize flow with a high kinetic
energy of 21 eV and the green ones with a low kinetic energy of 9 eV.
For Ar, the dominance of the external field over the dipole response
(mind the scaling in the left middle panel) enhances the ponderomotive
effects and thus the high-energy electrons can only escape after the
last peak, thus in forward direction. For Na$_9^+$, on the other
hand, the external field is considerably weaker which allows
high-energy electrons to escape already the second last peak pointing
in negative direction. This, however, should not be taken as general
rule. The subtle interplay of drivers in \NaCl inhibits simple
estimates.

\subsection{Impact of the systems geometry}

\begin{SCfigure}
 \includegraphics[width=0.5\linewidth]{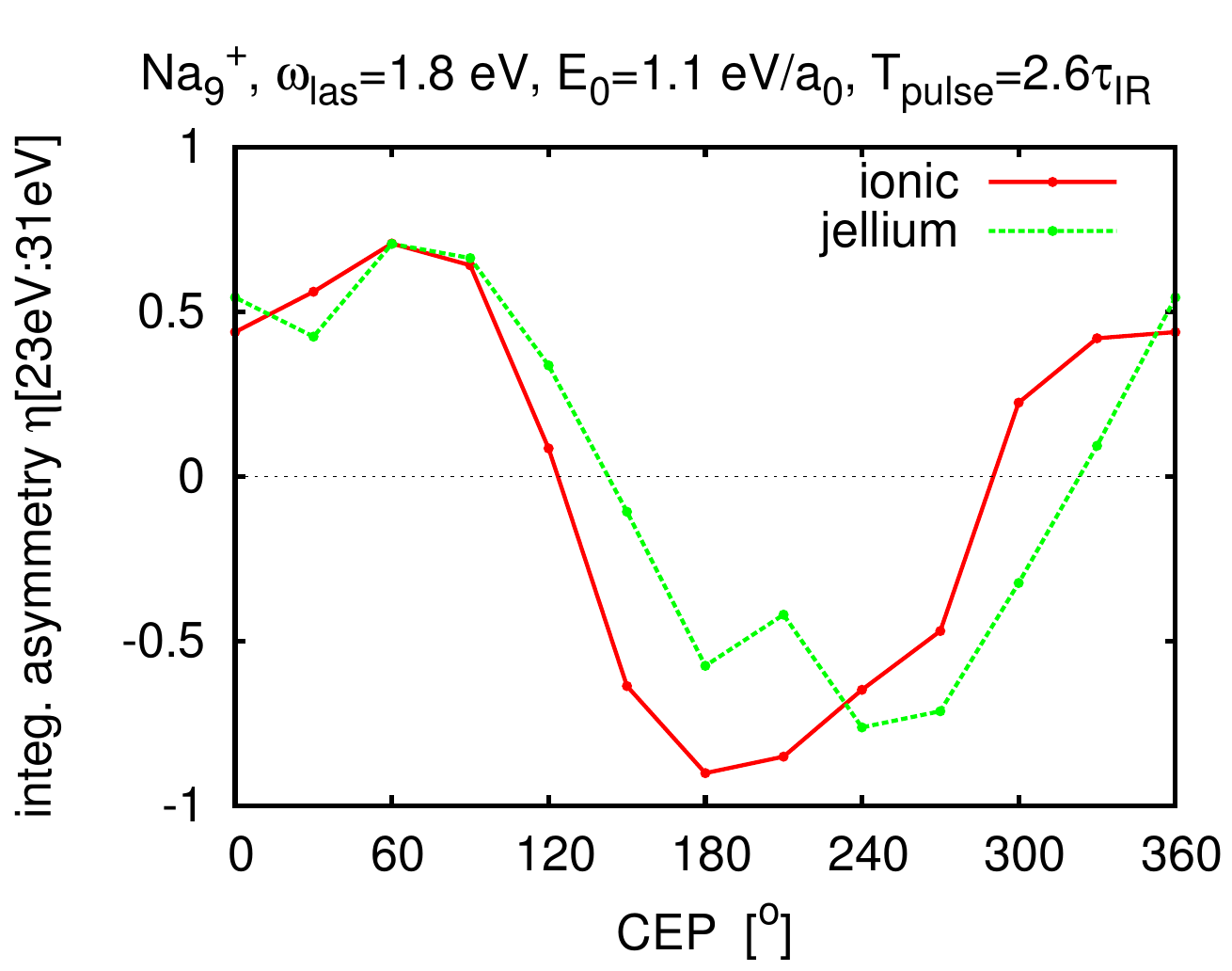}
 \caption{\label{fig:na9p-o13-compare} Integrated asymmetry
   $\eta[23eV:31eV]$ versus CEP comparing results from ionic
   background with jellium background.  Laser parameters were:
   $\omega_\mathrm{las}=1.8$ eV, $E_0=1.1$ eV/a$_0$, and 
   $T_\mathrm{pulse})=\tau_\mathrm{IR}$.  }
\end{SCfigure} 
To check the effect of ionic background, we compare in figure
\ref{fig:na9p-o13-compare} results from a calculation with ionic
background with one using a soft jellium background \cite{Rei03a}.
We use for the jellium parameters a Wigner-Seitz radius of
$r_s=3.65\bohr$ and surface thickness of $1\bohr$ to place the surface
plasmon resonance at similar spectral position as for the case of
ionic background.  The pulse length is chosen in the critical regime
to explore most sensitivity to system parameters. The results are
qualitatively the same. There are, of course, differences in
detail. But these could very well be due to the slightly different
spectral properties of the two systems. We can safely conclude that
details of the background structure do not matter much for the
structure of ARPES at these (low) laser frequencies. Note, however,
that a  sensitivity to ionic detail develops for laser pulses with
higher frequencies a,d was found in earlier studies on angular
distributions \cite{Poh04b}.

\subsection{Impact of laser parameters}

So far, we have considered laser pulses with frequency and intensity
close to the experimental conditions of former studies in C$_{60}$. It
is, of course, interesting to check the influence of laser
parameters. We do that here for the case of \NaCl.
\begin{figure}
 \centerline{
 \includegraphics[width=0.8\linewidth]{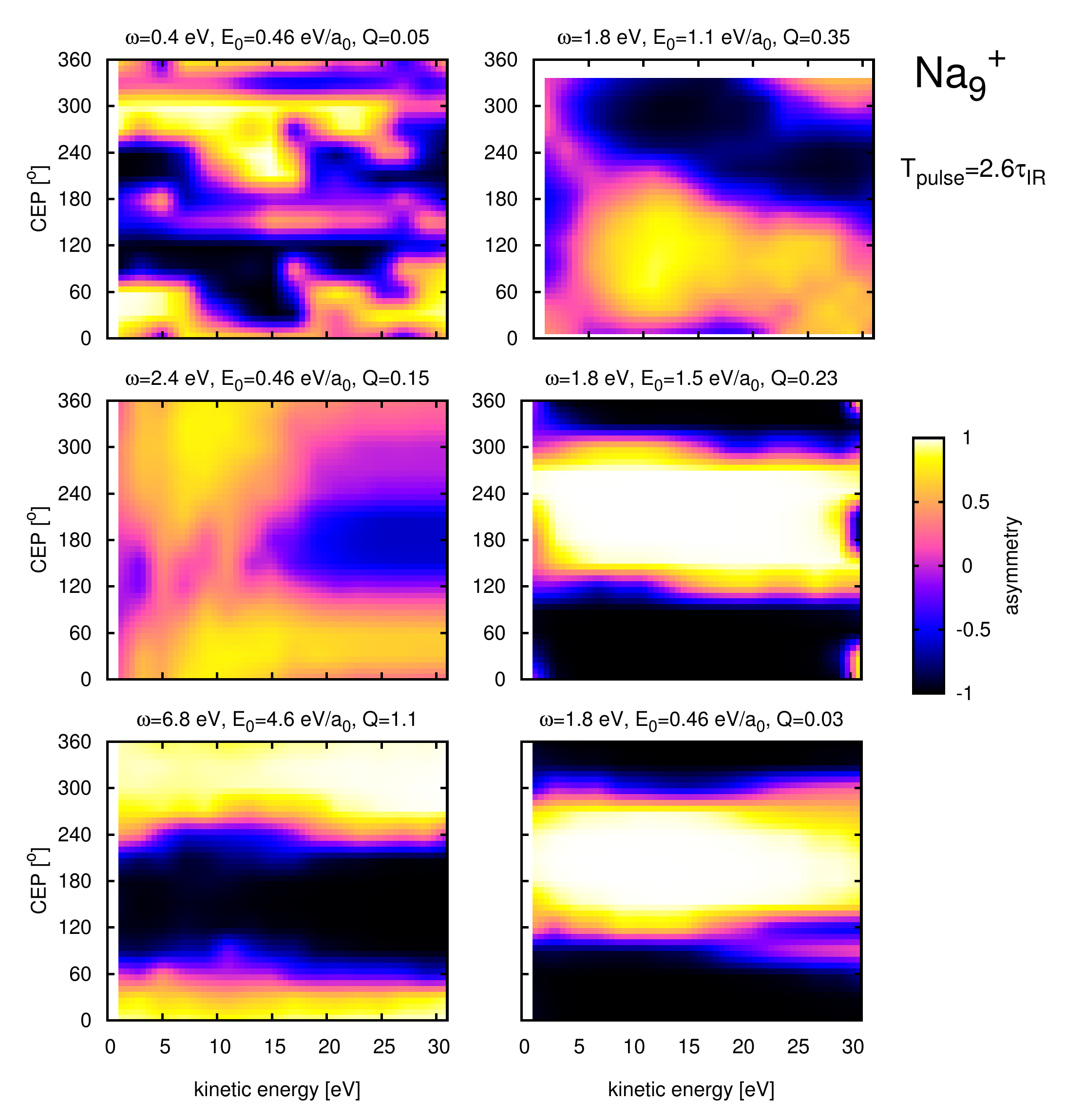}
}
 \caption{\label{fig:na9p-pesang-mixed} Asymmetry $\eta$ in the plane
   of kinetic energy and CEP for Na$_9^+$,
   $T_\mathrm{pulse}=2.6\tau_\mathrm{IR}$, and various laser parameters
   as indicated. Together with the laser parameters, we also
provide the total ionization $Q$ induced by laser excitation.
  }
\end{figure} 
Figure \ref{fig:na9p-pesang-mixed} shows the asymmetry $\eta$ in the
plane of kinetic energy and CEP for Na$_9^+$ and a variety of laser
frequencies and intensities.  The left panels collect variation of
frequency and the right panels variation of field strength (for fixed
frequency $\omega_\mathrm{las}=1.8$ eV).  Pulse length is
$T_\mathrm{pulse}=2.6\tau_\mathrm{IR}$, the same in all cases.  The
pattern differ dramatically for the different laser conditions.  Let
us try to give an interpretation to each case in the light of what we
have learned above.

The left column collects variation of frequency. The high-frequency
case (lower left panel) shows no energy dependence at all. This
reminds the field dominated case of short pulses in figure
\ref{fig:compare-asym-bins} and, in fact, we encounter here a similar
situation. At an absolute scale, the pulses are extremely short thus
overruling any system response.  The resonant case (middle left panel)
at surface plasmon frequency looks somewhat blurred resembling the
cases with longer pulses in figure
\ref{fig:compare-asym-bins}. Indeed, we deal effectively with a long
pulse because the resonant excitation induces dipole oscillations
which carry on long after the external field has been turned off.  The
extremely low frequency (upper left panel) moves the photon field well
out of resonance, similar as the IR field was off resonance for Ar.
But the pattern do not show the regular trends as in case of Ar.  The
quasi-static field at this extremely low frequency creates a very
special situation. Although field strength seems weak, the long time a
certain pulling force persists leads to large ponderomotive
excursions. This together with a small amount of comparatively  much
faster plasmon oscillations generates these sort of chaotic, quickly
changing asymmetries. 

The right column of figure \ref{fig:na9p-pesang-mixed} collects
variation of field strength. The upper right panel repeats the
reference case which we had studied before in detail. Lower field
strength (middle and lower right panels) reduce the ponderomotive
effects even further and so diminish the differences between
high-energy and low-energy electrons. What remains is pronounced
forward emission for CEP $0^{\circ}$ following the strongest peak in
that case and correspondingly pronounced backward emission for
$180^{\circ}$.

\section{Conclusion}

In continuation of a previous study on C$_{60}$, we have investigated
the system dependence of the impact of the carrier-envelope phase
(CEP) of few-cycle laser pulses using as two very different systems an
Ar atom and a Na$_{9}^+$ cluster. As theoretical tool for this survey
we use time-dependent density functional theory using standard local
energy-density functionals. A self-interaction correction is added to
achieve a correct dynamical description of emission properties.  As in
previous experimental and theoretical studies, we take as key
observable the detailed electronic emission properties in angular
resolved photo-electron spectra (ARPES). The angular information is
compressed to the asymmetry of the angular distribution as function of
energy which yields a compact measure well suited for a survey of
varying conditions.

The two systems under consideration differ dramatically in their
spectral properties. The Ar atom has a rather rigid dipole response
with the first excitations lying high above IR frequencies.  In this
respect, the Ar atom is similar to the C$_{60}$ cluster.  Although the
latter has a much more complex background structure the results for
asymmetry as function of CEP are practically the same for these both
systems.  Quite differently, the \NaCl cluster has a large
polarizability and its spectrum is dominated by the strong surface
plasmon resonance in the visible range, above IR frequencies but not
too far away. In consequence, the results for ARPES from few cycle
pulses differ dramatically from those of the Ar atom.  By looking at
the time evolution of density in detail, this difference could be
clearly related to the different dynamical polarizability. The dipole
moment of Ar follows the external field without delay whereas for \NaCl
it starts with a sizable phase shift from the onset and develops
further phase deviations in the course of dynamics. This produces a
much different interplay between electron emission, triggered by the
dipole moment, and subsequent ponderomotive motion, governed by the
external field. The more involved interplay between dipole and
external field in case of \NaCl hinders to develop simple
predictions. A full simulation taking into account properly the
dynamical response of a system is necessary. 

Concerning variation of laser parameters, the length of the pulse
plays a key role for the impact of the CEP. Longer pulses covering
several optical cycles render the CEP unimportant. Very short pulses,
up to 2 cycles, reduce the impact of the systems response and let the
effects of external field dominate. In between comes a transitional
region where the CEP dependence of ARPES changes quickly and becomes
extremely system dependent. We also checked the influence of the other
laser parameters for the more responsive system \NaCl in the critical,
transitional regime of pulse lengths and find, not surprisingly, a
sensitive dependence to any parameter.  The qualitative structure of
the results can be explained in each case by the relative importance
of ponderomotive motion in the external field and dipole response of
the system.

\bigskip

\noindent
Acknowledgment:\\ We thank the regional computing center of the
university Erlangen-N\"urnberg for generous supply of computer time
for the demanding calculations.  The work was supported by the
Institut Universitaire de France and a French ANR contract LASCAR
(ANR-13-BS04-0007).

\bigskip

\bibliographystyle{iopart-num}
\bibliography{na9p-CEP}

\end{document}